\documentclass[aps,pra,twocolumn,floatfix,10pt,nofootinbib]{revtex4}
\usepackage{amsmath,mathrsfs,graphicx}
\usepackage{multirow}
\begin{document}
\title[Absolute absorption on rubidium D$_{1}$ line]{Absolute absorption on rubidium D$_{1}$ line: including resonant dipole-dipole interactions}
\author{Lee Weller, Robert J Bettles, Paul Siddons, Charles S Adams and Ifan G Hughes}
\affiliation{Department of Physics, Durham University, South Road, Durham, DH1~3LE, UK}
\date{\today}

\begin{abstract}
\noindent 
Here we report on measurements of the absolute absorption spectra of dense rubidium vapour on the D$_{1}$ line in the weak-probe regime for temperatures up to 170~$^{\circ}$C and number densities up to $3~\times~10^{14}$~cm$^{-3}$.  In such vapours, modifications to the homogeneous linewidth of optical transitions arise due to dipole-dipole interactions between identical atoms, in superpositions of the ground and excited states.  Absolute absorption spectra were recorded with deviation of 0.1\% between experiment and a theory incorporating resonant dipole-dipole interactions.  The manifestation of dipole-dipole interactions is a self-broadening contribution to the homogeneous linewidth, which grows linearly with number density of atoms.  Analysis of the absolute absorption spectra allow us to ascertain the value of the self-broadening coefficient for the rubidium D$_{1}$ line: $\beta/2\pi$~=~$(0.69 \pm 0.04)$~$\times$~10$^{-7}$~Hz~cm$^{3}$, in excellent agreement with the theoretical prediction.
\end{abstract}
\maketitle
\section{Introduction}
\label{Introduction}
Understanding and controlling the propagation of light through a hot vapour is a burgeoning area of research.  Specific topics of interest include modifying the speed of light propagation in a medium (``slow light'')~\cite{boyd2002slow,fleischhauer2005electromagnetically}; studying the quantum interface between light and atomic ensembles~\cite{hammerer2010quantum}, for example in quantum memory~\cite{julsgaard2004experimental}; all-optical switching~\cite{PhysRevA.81.043838,dawes2005all} and measurements of transition amplitudes between low-lying states providing a valuable test of {\it ab initio} calculations~\cite{rafac1998measurement}.

An understanding of the absolute susceptibility of a hot vapour is useful for many applications, including: analysing electromagnetically induced transparency (EIT) spectra~\cite{badger2001hyperfine,mohapatra2007coherent}; performing far-off resonance Faraday spectroscopy~\cite{siddons2009gigahertz}; realising a broadband optical delay line~\cite{vanner2008broadband,camacho2007wide}; designing a Faraday dichroic beam splitter for Raman light~\cite{abel2009faraday}; generating frequency up-converted light~\cite{vernier2010enhanced}; primary spectroscopy~\cite{truong2011quantitative} and producing suitable spectral features for laser stabilisation~\cite{marchant2011off,kemp2011analytical}.

In a previous paper~\cite{siddons2008absolute} we developed a model for the electric susceptibility that allowed us to make quantitative predictions for the absorptive and dispersive properties of hot rubidium (Rb) vapour probed in the vicinity of the D lines\footnote{For an alkali-metal atom the D$_{1}$ transition is $n^{2}$S$_{1/2}$~$\rightarrow$~$n^{2}$P$_{1/2}$, where $n$ is the principal quantum number of the valence electron, and the D$_{2}$ transition is $n^{2}$S$_{1/2}$~$\rightarrow$~$n^{2}$P$_{3/2}$.}.  Measurements of the frequency dependence of the absolute Doppler-broadened absorption coefficient were performed for temperatures up to 36.6~$^{\circ}$C and showed excellent agreement with the theoretical model.  Resonant dipole-dipole interactions among Rb atoms were not included in the model, as they do not become important unless the vapour has a temperature exceeding $\sim125~^{\circ}$C.  However, there is a strong motivation to working at higher temperature as in this case high optical depths are accessible even in cells with lengths of only a few millimetres or a few microns~\cite{James2011Opacity}.  Here we show that by including self broadening in the model for susceptibility, experiment and theory agree to within 0.1$\%$ up to densities of $3~\times~10^{14}$~cm$^{-3}$.  At these higher densities the resonant dipole-dipole interactions between two identical atoms, in superpositions of the ground and excited states, gives rise to the phenomenon of self-broadening~\cite{gorris1997doppler}.  The modification of the medium's susceptibility as a consequence of the dipole-dipole interactions is the subject of this paper.  In this work we extend the theoretical and experimental study of the absolute absorption spectroscopy of Rb on the D$_{1}$ line for temperatures up to $\sim170~^{\circ}$C, corresponding to a number density four orders of magnitude larger than in~\cite{siddons2008absolute}.  

The structure of the remainder of the paper is as follows.  In section $\ref{Theoretical Considerations}$ we give an overview of the theoretical calculations behind self-broadening and relate them to the previously measured values for alkali-metal atoms;  in section $\ref{Experimental Method}$ we describe the experimental apparatus, giving the results of our investigation in section $\ref{Results and Discussions}$, before drawing our conclusions in section $\ref{Conclusions}$.
\section{Theoretical Considerations}
\label{Theoretical Considerations}
The susceptibility we developed is calculated by summing the contributions from all dipole-allowed transitions~\cite{siddons2008absolute}.  Each transition has a Voigt profile with two contributions: a homogeneous Lorentzian lineshape arising from spontaneous emission from the excited state (with natural linewidth $\Gamma_\mathrm{0}$), and an inhomogeneous Gaussian lineshape to account for the atomic motion.  At the higher densities considered here there is a significant contribution to the total Lorentzian width, $\Gamma_\mathrm{tot}$, from the interaction among the atoms:
\begin{eqnarray}
\Gamma_\mathrm{tot} = \Gamma_\mathrm{0} + \Gamma_\mathrm{self} = \Gamma_\mathrm{0} + \beta \mathcal{N},
\label{spectralfit}  
\end{eqnarray}
where $\beta$ is the self-broadening coefficient.  The dominant effect arises from the dipole-dipole interaction between two identical atoms, in superpositions of the ground and excited states.  The binary dipole-dipole approximation remains valid for densities up to $\mathcal{N}r_\mathrm{w}^{3}\ll1$ where $r_\mathrm{w}\,=\,\sqrt{\,\beta\,/\,2v_{0}\,}$~\cite{sautenkov1996dipole} and $v_{0}$ is the most probable atomic velocity in the vapour.  This gives $\mathcal{N}\geq10^{17}$~cm$^{-3}$ (corresponding to a temperature $\sim360~^{\circ}$C) which is valid for this work.    

The impact regime is defined by the inequality $\vert\Delta\vert<\omega_\mathrm{w}$, where $\Delta$ is the detuning of the laser light from resonance, ($\Delta\,=\,\omega_{\mathrm{L}}\,-\,\omega_{0}$, with $\omega_{\mathrm{L}}$ the laser angular frequency, and $\omega_{0}$ the angular frequency of the resonance transition), and $\omega_\mathrm{w}$ is the Weisskopf angular frequency.  For the Rb D$_{1}$ line the Weisskopf angular frequency defined by $\omega_\mathrm{w}\,\approx\,2\pi\,\times\,\sqrt{\,v_{0}^{3}\,/\,3\beta\,}$~\cite{PhysRevA.27.1851}, is approximately $2\pi \,\times\,$4~GHz for the temperatures studied here.  The detunings from resonance accessible with our apparatus are less than the Weisskopf frequency, therefore we need to evaluate the self-broadening parameter $\beta$ in the impact regime within the binary-collision approximation.   

In a comprehensive treatment of collisional line broadening Lewis~\cite{lewis1980collisional} provides an expression for the self-broadening parameter for alkali-metal atoms in the binary-collision approximation, which is $\beta\,=\,2\,f\,c\,r_{0}\,\lambda\,\sqrt{g_{g}/g_{e}}$.  Here, $f$ is the absorption oscillator strength for the transition; $c$ is the speed of light; $r_{0}$ is the classical radius of the electron; $\lambda$ is the resonance transition wavelength; $g_{g}$ and $g_{e}$ are the degeneracies of the ground and excited states, respectively.  To highlight the physical mechanism underpinning the interaction we rewrite the formula for the self-broadening coefficient as
\begin{eqnarray}
\beta_{1} &=& \frac{2}{9 \hbar \epsilon_{0}} d^{2}_{1} = 2 \pi \times \Gamma_{1} \left(\frac{\lambda_{1}}{2 \pi}\right)^{3}, \label{firstequation}\\
\beta_{2} &=&\sqrt{2} \frac{2}{9 \hbar \epsilon_{0}} d^{2}_{2} = 2 \pi \times \sqrt{2}~\Gamma_{2} \left(\frac{\lambda_{2}}{2 \pi}\right)^{3},
\label{secondequation} 
\end{eqnarray}
where $d_{1},d_{2}$, $\Gamma_{1},\Gamma_{2}$ and $\lambda_{1},\lambda_{2}$, are the reduced dipole matrix elements, natural linewidths and wavelengths for the D$_{1}$ and D$_{2}$ lines, respectively, for alkali-metal atoms.  The $d^{2}$ terms in these equations highlight the dipole-dipole origin of the self-broadening interaction.  The dependence on $\Gamma$ demonstrates that the lines broaden by an amount equal to the natural-broadening per atom within a volume equal to the reduced wavelength cubed.  Note also that the self-broadening coefficient is $\sqrt{2}$ larger for the D$_{2}$ line compared to the D$_{1}$ line.

In Table~\ref{firsttable} we have collated experimental measurements of the D$_{1}$ and D$_{2}$ self-broadening coefficients for Na, Rb and Cs, and compared these measurements with the theoretical predictions of equations (\ref{firstequation}) and (\ref{secondequation}).  A variety of experimental techniques were used for the measurements (line-wing absorption, and reflection spectroscopy); all of the values are in excellent agreement with the theoretical prediction within the quoted uncertainty.

\begin{table}[tb]
\centering
\caption{Theoretical and measured values of the impact self-broadening coefficients $\beta_{1,2}/2\pi$ for Na, Rb and Cs on the D$_{1}$ and D$_{2}$ lines. Theoretical predictions are from equations (\ref{firstequation}) and (\ref{secondequation}).}
\begin{tabular}{ccccc}
\hline
& \multicolumn{2}{c}{D$_{1}\;\textnormal{in}\; 10^{-7}~\textnormal{Hz~cm}^3$}   & \multicolumn{2}{c}{D$_{2}\;\textnormal{in}\; 10^{-7}~\textnormal{Hz~cm}^3$}   \\
Element   & Theory      & Measured  				                        			& Theory       & Measured												          			\\\hline	  											
Na				& 0.51		    &	0.49$\pm$0.07 \cite{PhysRevA.27.1851}						& 0.72 				 & 0.74$\pm$0.11 \cite{PhysRevA.27.1851}				  \\
Rb				& 0.73 			  &	0.69$\pm$0.04 present	  												& 1.03  			 & 1.10$\pm$0.17 \cite{kondo2006shift}						\\
Cs				& 0.83		    &	0.75$\pm$0.11 \cite{vuletic1993measurement}		 	& 1.16      	 & 1.15$\pm$0.23 \cite{akulshin1982collisional}	  \\\hline
\label{firsttable}
\end{tabular}
\end{table}

\begin{figure}[tb]
\centering
\includegraphics*[width=0.45\textwidth]{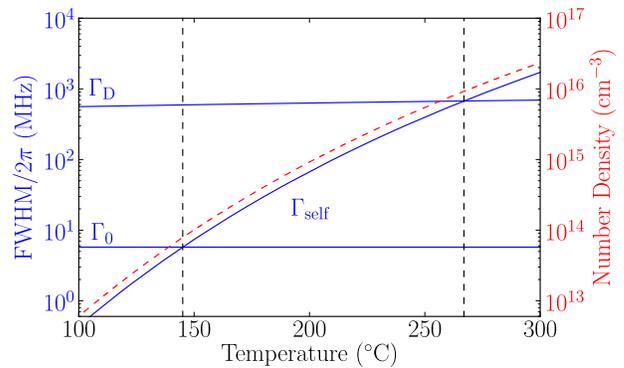}
\caption{The solid blue lines show natural, $\Gamma_\mathrm{0}$, Doppler, $\Gamma_\mathrm{D}$, and self-broadened, $\Gamma_\mathrm{self}$, FWHM for the $^{87}$Rb D$_{1}$ line as a function of temperature.  The dotted red line shows number density as a function of temperature.  The dotted black lines define the regimes where $\Gamma_\mathrm{D}>\Gamma_\mathrm{0}>\Gamma_\mathrm{self}$ (leftmost region), $\Gamma_\mathrm{D}>\Gamma_\mathrm{self}>\Gamma_\mathrm{0}$ (central region), and 
$\Gamma_\mathrm{self}>\Gamma_\mathrm{D}>\Gamma_\mathrm{0}$ (rightmost region).}
\label{Figure1}
\end{figure}

To illustrate how the density maps on to the vapour temperature, Figure $\ref{Figure1}$ shows the natural, $\Gamma_\mathrm{0}$, Doppler, $\Gamma_\mathrm{D}$, and self-broadened, $\Gamma_\mathrm{self}$, full-widths half-maximum (FWHM) for the $^{87}$Rb D$_{1}$ line as a function of temperature.  The Rb number density~\cite{alcock1984vapour} is also plotted.  For temperatures below $\sim145~^{\circ}$C the Doppler width is two orders of magnitude larger than the homogeneous linewidth, and natural-broadening exceeds self-broadening; for temperatures above $\sim265~^{\circ}$C self-broadening dominates; and in the intermediate regime self-broadening dictates the inhomogeneous width.  The optical depth on resonance increases rapidly with temperature, therefore for experiments where it is desirable for a large fraction of the light to be transmitted this necessitates working far from  resonance.  In the range of temperatures to which we gain access in the experiments we report here Doppler-broadening exceeds the homogeneous (natural- and self-) broadening; however, in the wing of the absorption line the profile is expected to have a Lorentzian profile~\cite{siddons2009off}. 

\section{Experimental Method}
\label{Experimental Method}
\begin{figure}[tb]
\centering
\includegraphics*[width=0.45\textwidth]{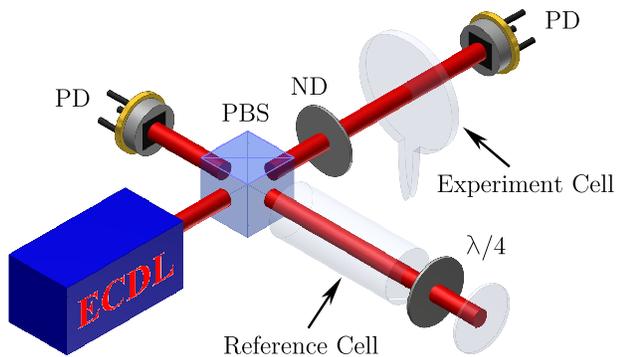}
\caption{(Colour online) Schematic of the experimental apparatus.  A beam passes through a polarisation beam splitter (PBS), providing linearly polarised light.  A small fraction of the beam is used to perform sub-Doppler spectroscopy in a room-temperature reference cell.  The probe beam is attenuated with a neutral-density filter (ND) before passing through a heated vapor cell before being collected on a photodiode (PD).}
\label{Figure2}
\end{figure}

Figure $\ref{Figure2}$ shows a schematic of the experimental apparatus used to observe the resonance line self-broadening on the D$_{1}$ transition of Rb.  An external cavity diode laser system (ECDL) with a wavelength of 795~nm was used for these measurements.  The beam had 1/e$^{2}$ radius $0.77\,\pm\,0.02$~mm and passed through a polarisation beam splitter providing linearly polarised light.  After attenuation by a neutral-density filter the beam was sent through a 2~mm heated vapour cell containing Rb in its natural abundance (72\% $^{85}$Rb, 28\% $^{87}$Rb).  The cell was placed in an oven which has two sections, one containing the body of the cell, the other the metal reservoir.  The oven is made from non-magnetic stainless steel and polyether ether ketone (PEEK); the latter is used owing to its heat-insulating properties.  The two oven sections are resistively heated independently.  In this investigation a constant temperature difference of $\sim5~^{\circ}$C was maintained  between the body of the cell and the reservoir; this is to prevent Rb from condensing on the windows.  The body and reservoir temperatures were measured with thermocouples.  After traversing the experiment cell the light impinges on a calibrated photodiode.  The frequency scan was linearised with a Fabry-Perot etalon (not shown), and calibrated by the use of hyperfine/saturated absorption spectroscopy~\cite{macadam1992narrow,smith2004hyperfine} in a natural-abundant room-temperature reference cell. 

To obtain good agreement between theory and experiment it is important that the atoms in the cell traversing the laser beam do not undergo hyperfine pumping into the other ground term hyperfine level.  This is achieved by working with a probe beam power much less than 100~nW~\cite{siddons2008absolute,sherlock2009weak}.  We measured transmission spectra for a range of temperatures from room temperature up to $\sim170~^{\circ}$C.  For each temperature five spectra were recorded and analysed. 

\section{Results and Discussions}
\label{Results and Discussions}
\begin{figure}[tb]
\centering
\includegraphics*[width=0.45\textwidth]{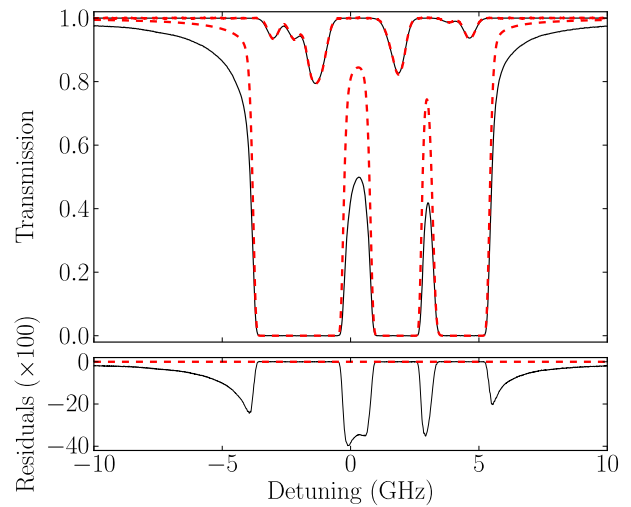}
\caption{Transmission plot for the comparison between experiment and theory for the Rb D$_{1}$ line, through a 2~mm vapour cell as a function of linear detuning, $\Delta/2\pi$.  The zero of detuning is taken to be the centre-of-mass frequency of the transition.  Solid black and dotted red lines show measured and expected transmission, respectively, for a temperature of 55~$^\circ$C and 170~$^\circ$C with $\Gamma_0/2\pi$~=~$5.746$~MHz~\cite{volz1996precision}.  Below the main figure is a plot of the residuals (the difference in transmission between theory and experiment) for the higher temperature.  The structure and magnitude of the residuals confirm how poor a fit the theory is to the data.}
\label{Figure3} 
\end{figure}

Figure $\ref{Figure3}$ shows a plot of the transmission of the Rb D$_{1}$ line versus linear detuning $\Delta/2\pi$.  The zero of the detuning axis is taken to be the centre of mass frequency of the transition in the absence of hyperfine splitting, taking into account the natural abundance of each isotope.  The solid black and dotted red lines show the measured and theoretical transmission, respectively, using the susceptibility of~\cite{siddons2008absolute}.  The two theory curves are generated assuming that the Lorentzian width is $\Gamma_0$, the natural width and with temperatures 55~$^\circ$C and 170~$^\circ$C in agreement with thermocouple measurements.  There is excellent agreement between theory and experiment for the lower temperature; however, at the higher temperature agreement is poor.  The discrepancy at the higher temperature is not surprising when recalling Figure $\ref{Figure1}$; self-broadening is expected to provide the dominant contribution to the homogeneous width in this regime.
       
\begin{figure}[tb]
\centering
\includegraphics*[width=0.45\textwidth]{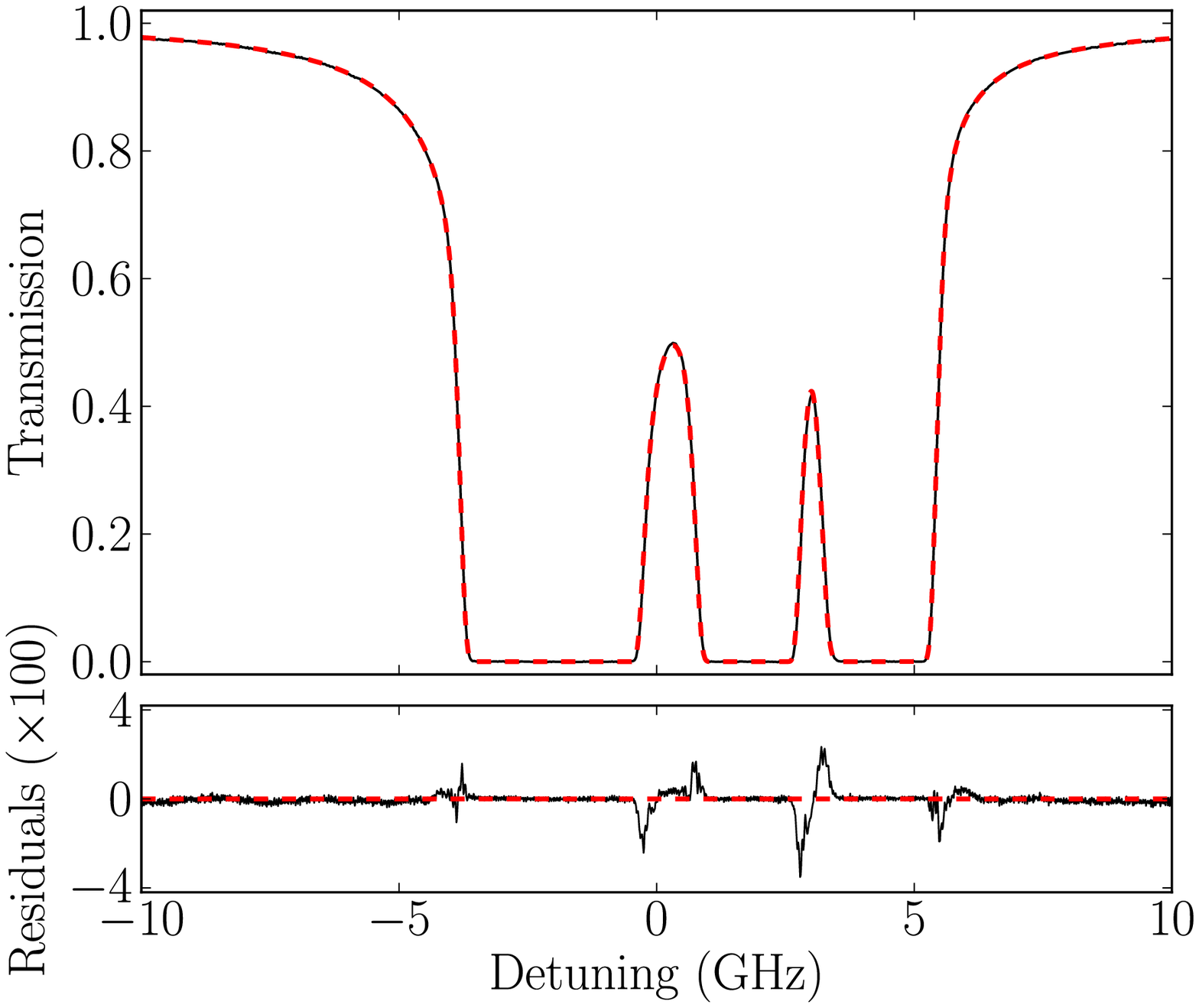}
\caption{Transmission plot for the comparison between experiment and theory for the Rb D$_{1}$ line, through a 2~mm vapour cell as a function of linear detuning, $\Delta/2\pi$.  The zero of detuning is taken to be the centre-of-mass frequency of the transition.  Solid black and dotted red lines show measured and expected transmission, respectively, for a temperature of (171.8 $\pm$ 0.2)~$^\circ$C with $\Gamma_\mathrm{tot}/2\pi$~=~(23.7 $\pm$ 0.2)~MHz.  Below the main figure is a plot of the difference in transmission between theory and experiment.  There is excellent agreement between model and data, with the exception of a small number of glitches where the linearisation of the laser scan was not adequate.}
\label{Figure4} 
\end{figure}

Based on the discussion in section \ref{Theoretical Considerations} we modify the form of the susceptibility in our theoretical model.  We expect the total Lorentzian linewidth to be a function of density, and hence temperature.  Both isotopes have two values of $F$ in the ground $^{2}$S$_{1/2}$ term ($^{85}$Rb has $F$ = 2 and 3; $^{87}$Rb has $F$ = 1 and 2).  The susceptibilities for transitions from each $F$ are allowed to have variable Lorentzian linewidths.  Figure $\ref{Figure4}$ shows a comparison of the data and modified theoretical prediction.  A least-squares fit\footnote{The Marquardt-Levenberg method~\cite{MATU} was used to perform a least-squares fit and extract the optimised parameters and their uncertainties.} of the data to the modified theory allows the total homogeneous linewidths, $\Gamma_\mathrm{tot}$, and temperature to be determined from the spectrum.  The agreement between theory and experiment is excellent.  There are minor glitches in the residuals where the transmission varies most rapidly - this is a manifestation of the imperfect linearisation of the laser scan.  Ignoring those anomalies, the deviation between theory and experiment is at the 0.1\% rms level.  Five spectra were taken for each temperature value, and the statistical uncertainty in the twenty measurements of the total homogeneous linewidth, $\Gamma_\mathrm{tot}$, was calculated as the standard error.  

\begin{figure}[tb]
\centering
\includegraphics*[width=0.45\textwidth]{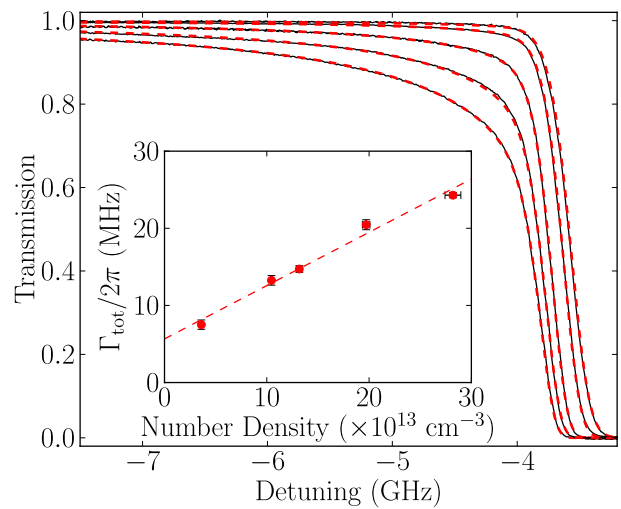}
\caption{Transmission plots for the comparison between experiment and theory for the $F = 2 \rightarrow F' = 1,2$ for $^{87}$Rb, through a 2~mm natural abundant vapour cell, as a function of linear detuning, $\Delta/2\pi$.  Solid black and dotted red lines show measured and expected transmission for several different number densities.  The insert shows total homogeneous linewidth versus number density.  The dotted red line shows a linear relationship between the two axes, with a gradient of  $\beta/2\pi$~=~(0.69 $\pm$ 0.04)~$\times$~10$^{-7}$~Hz~cm$^{3}$ and an intercept of $\Gamma_0/2\pi$~=~(5.7 $\pm$ 0.7)~MHz.} 
\label{Figure5}
\end{figure}

In Figure $\ref{Figure5}$ the transmission as a function of linear detuning through a 2~mm natural abundant vapour cell is measured in the red-detuned wing for several different temperatures.  Despite being in the regime where the Doppler width exceeds the homogeneous width, as discussed above, the absorption profile far from resonance has a Lorentzian profile.  Measuring the total homogeneous width as a function of temperature allows us to evaluate the self-broadening coefficient.  For each temperature, the Rb number density is deduced from~\cite{alcock1984vapour}, and the total homogeneous linewidth and its uncertainty are extracted from the least-squares fit.  For the density range investigated the inset to Figure $\ref{Figure5}$ shows that the spectral width $\Gamma_\mathrm{tot}$ is linear in number density, as expected from equation~\ref{spectralfit}.  From the slope we find a value for the self-broadening coefficient for the Rb D$_{1}$ line $\beta/2\pi$~=~$(0.69 \pm 0.04)$~$\times$ 10$^{-7}$~Hz~cm$^{3}$, and infer a natural linewidth $\Gamma_0/2\pi$~=~(5.7 $\pm$ 0.7)~MHz from the intercept of the fit.  The experimentally determined self-broadening coefficient is in excellent agreement with the theoretical prediction presented in Table~\ref{firsttable}. 

The statistical variation in the Rb density evaluated from the five spectra in each series of measurements is very small (0.6\%); two possible systematic errors in the evaluation of number density are the accuracy of the vapour pressure formula in~\cite{alcock1984vapour} and the small difference between the temperature of the atoms interacting with the light and the atoms in the metal reservoir.  The centre of the Rb absorption lines are shifted in dense vapour owing to local-field effects~\cite{maki1991linear}.  The shift is expected to be proportional to the atomic density~\cite{sautenkov1996dipole}, and for the highest density studied in this work is of the order of 3~MHz - too small to be evident.  Note that recently it has been shown that at densities of the order of $10^{16}$~cm$^{-3}$ dipole-dipole interactions can lead to the saturation of the resonance susceptibility~\cite{James2011Opacity}. 
\section{Conclusions}
\label{Conclusions}
We have experimentally shown that for hot Rb vapour with density up to $3~\times~10^{14}$~cm$^{-3}$ the Lorentzian component of the D$_{1}$ line is modified as the result of dipole-dipole interactions.  The off-resonant absorption lineshape is sensitive to the total homogeneous linewidth, which grows linearly with the number density of atoms.  We have shown that a simple modification of the theoretical model for electric susceptibility to take into account self-broadening leads to excellent agreement between theory and experiment.  Absolute absorption spectra have been measured for detunings of the order of the Weisskopf frequency or less and for temperatures up to $\sim170~^{\circ}$C with deviation of 0.1\% between theory and experiment.  These measurements allowed us to ascertain the value of the self-broadening coefficient for the Rb D$_{1}$ line to be $\beta/2\pi$~=~$(0.69 \pm 0.04)$~$\times$~10$^{-7}$~Hz~cm$^{3}$, which is in excellent agreement with the theoretical impact self-broadening coefficient.  Our results give new insight into the resonant dipole-dipole collision physics in dense atomic vapours.
\section*{Acknowledgements}
This work is supported by ESPRC.  Thanks to Mr C. Jolly for funding Robert J. Bettles.  We thank James Keaveney for discussion and theoretical assistance with the numerical fitting programme.  
\bibliographystyle{unsrt}
\bibliography{myreferences}
\end{document}